\documentclass[%
aps,%
amsmath,amssymb,floatfix,
reprint,%
pre,%
]{revtex4-1}

\usepackage{graphicx}
\usepackage{dcolumn}
\usepackage{bm}
\usepackage{amssymb}
\usepackage{mathtools}
\usepackage{multirow}
 \usepackage[table,xcdraw]{xcolor}
\usepackage{hhline}
 \usepackage{hyperref}

\newcommand{\1}{\begin{equation}}
\newcommand{\2}{\end{equation}}
\newcommand{\ea}{\begin{eqnarray}}
\newcommand{\ee}{\end{eqnarray}}

\newcommand{\RR}{\mathbf{r}}
\newcommand{\VV}{\dot{\mathbf{r}}}

\begin{document}


\title{Active Ornstein-Uhlenbeck model for self-propelled particles with inertia}



 \author{G.\ H.\ Philipp Nguyen, Ren\'e Wittmann, Hartmut L\"owen}
 \email[]{hlowen@hhu.de}
 \affiliation{Institut f\"{u}r Theoretische Physik II: Weiche Materie, Heinrich-Heine-Universit\"{a}t D\"{u}sseldorf, D-40225 D\"{u}sseldorf, Germany}


\date{\today}

\begin{abstract}
Self-propelled particles, which convert energy into mechanical motion, exhibit inertia if they have a macroscopic size or move inside a gaseous medium, in contrast to micron-sized overdamped particles immersed in a viscous fluid. 
Here we study an extension of the active Ornstein-Uhlenbeck model, in which self-propulsion is described by colored noise, to access these inertial effects. 
We summarize and discuss analytical solutions of the particle's mean-squared displacement and velocity autocorrelation function for several settings 
ranging from a free particle to various external influences, like a linear or harmonic potential and coupling to another particle via a harmonic spring.
 Taking into account the particular role of the initial particle velocity in a nonstationary setup, we observe all dynamical exponents between  zero and four.
After the typical intertial time, determined by the particle's mass, the results inherently revert to the behavior of an overdamped particle
with the exception of the harmonically confined systems, in which the overall displacement is enhanced by inertia.
We further consider an underdamped model for an active particle with a time-dependent mass, which critically affects the displacement in the intermediate time-regime.
Most strikingly, for a sufficiently large rate of mass accumulation, the particle's motion is completely governed by inertial effects as it remains superdiffusive for all times.
\end{abstract}

\maketitle

\section{Introduction}\label{ra_sec1}

The physics of self-propelled particles is a flourishing research arena.
There exist many different biological microswimmers in nature, for instance, bacteria and unicellular protozoa, which typically generate their swimming motion with flagella or cilia powered by molecular motors~\cite{ecoli, protozoa}.
Janus particles are examples of synthetic microswimmers, which possess surfaces with two distinct physical or chemical properties.
This asymmetric structure leads to self-propulsion via various mechanisms~\cite{janus}.
Even on the single particle level,
active motion is a nonequilibrium phenomenon, therefore challenging a basic modeling from a statistical mechanics point of view.
In the last decades, various simple models were designed and proposed for single active particles including
self-propulsion generated by nonlinear friction \cite{Ebeling,Romanchuk_review}, 
 by non-reciprocal bead motions \cite{Golestanian_Najafi}, and by an internal driving force combined 
with overdamped orientational Brownian dynamics
\cite{GolestanianPRL2007,tenHagen2009,tenHagen_JPCM_2011}, the latter leading to the standard model of active Brownian particles (ABPs) \cite{review}.

More recently, the maybe simplest nontrivial model for an overdamped fluctuating self-propelled particle in a viscous fluid was proposed.
Such an active Ornstein-Uhlenbeck particle (AOUP) possesses a stochastic driving force whose memory decays exponentially in time, leading to a persistence in the particle motion which mimicks the activity. 
This model, originally proposed by Ornstein and Uhlenbeck to study velocity distributions of passive particles \cite{uhlenbeck1930theory} 
and subsequently exploited for various other physical and mathematical problems \cite{OUP_applications0,OUP_applications1,OUP_applications2,luczka2005}, 
has by now become a basic reference for active motion \cite{AOUP_MIPS,AOUP_application1,sandford2018,solon2015,AOUP_general1,dabelow2019,fily2019,caprini2019entropyPROD,caprini2020velocityCORR1d,caprini2021FDR,AOUP_general2,chains,AOUPreview}. 
Although the AOUP model does not resolve the orientational degrees of freedom, it admits some characteristic features of activity, like persistent motion, surface accumulation and, most prominently, motility-induced phase separation (MIPS)~\cite{AOUP_MIPS,Tailleur_review}.
Describing self-propelled motion by an AOUP has the advantage that exact analytical solutions can be obtained for a large range of problems~\cite{harmonic1, harmonic2, sandford2017,AOUP_analytical1, AOUP_analytical2, Caprini_2018,Caprini2020}.
Moreover, the model provides a convenient basis to develop the theoretical description of more complex settings of interacting particles \cite{marconimaggi2015,farage2015,AOUP_application2, AOUP_application3, AOUP_application4,AOUPmodel,AOUPmodel2,capriniSM2018,wittmann2019}.
The experimental relevance of the AOUPs model has been also demonstrated
for a passive tracer particle in an active bath \cite{Maggi2014expAOUP, Maggi2017expAOUP}.

If the self-propelled object has a macroscopic size or moves in a gaseous medium, the emerging inertial effects pose some new challenges for theoretical modeling.
Depending on whether the motion is in a gas or a viscous medium, this underdamped active matter can be divided into two classes, namely ''dry'' and ''wet'' systems.
Wet particles are affected by hydrodynamic effects, described within the Navier-Stokes equations~\cite{Klotsa2019}, where the probably most prominent example from nature is a school of fish.
In contrast, dry particles only perform a practically undamped motion due to their inertia.
Apart from nature's typical realization of such a system in a flock of birds,
there is a large range of dry inertial particles whose motion is still affected by fluctuating random kicks of the surrounding medium.
Whirling fruits self-propelling in the air~\cite{RabaultFC2019} and small animals such as insects~\cite{Mukundarajan2016,beetlePREPRINT} are macroscopic examples found in nature. 
Besides these biological organisms, there are also artificial dry self-propelled particles.
Mesoscopic dust particles in plasmas, the so-called ''complex plasma'', can be brought into a joint underdamped self-propulsion by nonreciprocal interactions~\cite{MorfillI2009, plasma2, IvlevBHDNL2015}
or photophoresis \cite{NosenkoLKRZT2020}.
Other examples of inertial dry active matter are man-made macroscopic granules self-propelling on a vibrating plate~\cite{NarayanRM2007,scholz2018inertial} 
or equipped with an internal vibration motor~\cite{DauchotD2019, LeoniPEENASA2020} and mini-robots~\cite{bots1, bots2}.
These various experimental realizations
have also triggered an increasing number of theoretical work \cite{scholz2018inertial,debnath2020,breoni2020,rocket,sandoval_inertia1,sandoval_inertia2,caprini_inertialABPSvelocity2021,omar2021_inertiaMIPS} 
considering dry active particles with inertia, see \cite{lowen2020inertial} for a recent review.

\begin{figure*} [t] \centering\hfill
\includegraphics[width=\textwidth] {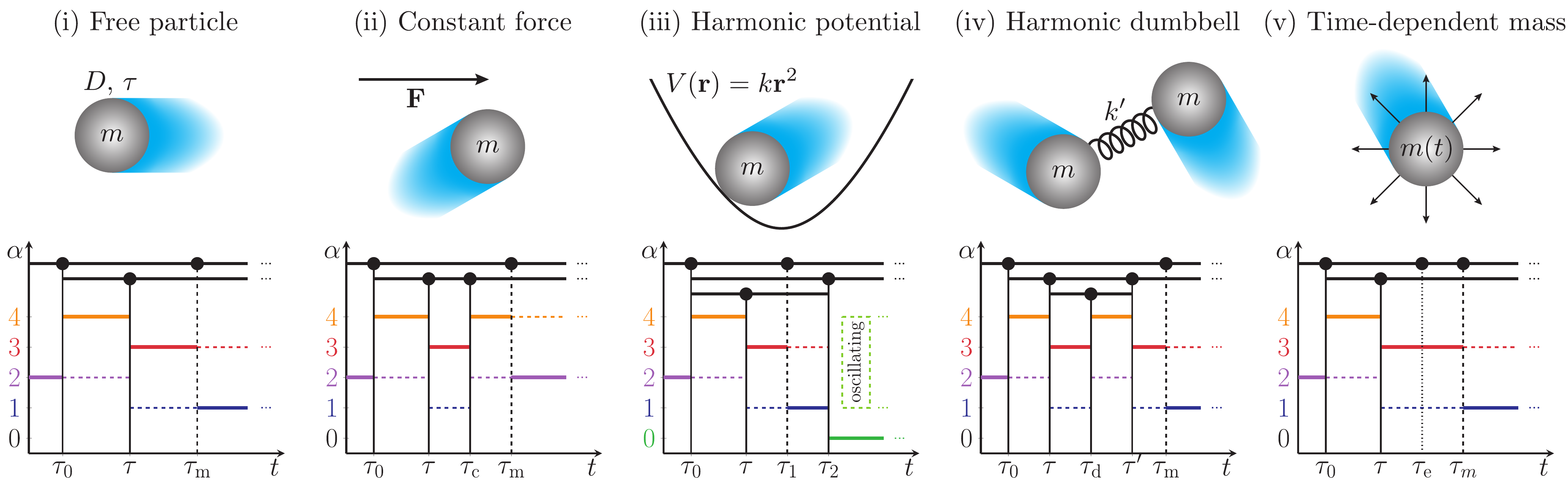}
\caption{
Overview of the main results.
\textbf{Top row:} schematic illustration of an inertial active Ornstein-Uhlenbeck particle (AOUP) at position $\RR(t)$ (gray sphere) and moving with velocity $\VV(t)$ (direction of the blue cloud).
Its stochastic motion, determined by Eq.~\eqref{eq_langevin}, depends on the initial conditions $\RR_0:=\RR(0)$ and $\VV_0:=\VV(0)$, its mass $m$, diffusion coefficient $D$, persistence time $\tau$ and the particular setup (i-v).
\textbf{Bottom row:} qualitative illustration of the observed dynamical exponents $\alpha(t)$ (colored lines) in the mean-squared displacement (MSD). %
The relevant time regimes are drawn in different layers (upper horizontal bars).
In each layer, the characteristic time scales (vertical bars with ticks on the $t$ axis) separating different regimes can be shifted horizontally (corresponding to a change of parameters), 
but their order is fixed (the big dots cannot get past each other). 
 Shifting a solid vertical bar prolongs one adjacent time regime and shortens the other one or even completely overlays the regime(s) from the layer(s) below.
The dashed vertical bar indicates the end of the inertial regime, which generally results in a fundamental change of the dynamical behavior. %
The exponents valid for the shown order of time scales are drawn as solid lines while the dotted lines become valid instead if the dashed vertical bar is shifted.
The dotted vertical line indicates a transition between two distinct regimes with the same exponent.
 The annotated time scales correspond to the shown setting, while their full definition and meaning %
is explained in the text for each scenario.
A detailed example of how to read this exponent diagram is given for a free particle in Sec.~\ref{sec_free3}.
\textbf{Columns:}
(i) force-free AOUP, cf.~Sec.~\ref{sec_free},
(ii) constant external force $\mathbf{F}$, cf.~Sec.~\ref{sec_const},
(iii) harmonic external potential with constant $k$ ($\alpha(t)$ is illustrated here for a spring with $k>0$), cf.~Sec.~\ref{sec_harm},
(iv) two harmonically coupled AOUPs 
with equal mass $m$ but different diffusion coefficient $D'$ and persistence time $\tau'$
($\alpha(t)$ is illustrated here for the center-of-mass coordinate $\mathbf{R}$), cf.~Sec.~\ref{sec_dumb},
and (v) with time-dependent mass $m(t)$ of constant slope $\dot{m}$ ($\alpha(t)$ is illustrated here for $\dot{m}<0$), cf.~Sec.~\ref{sec_mass}.
}
\label{fig_illustration}
\end{figure*}

In this paper, we study in detail the dynamical properties of an AOUP, 
whose translational motion is affected by inertia \cite{inertial0,inertial}.
Our motivation for this choice is twofold.
First, providing the simplest description of activity subject to inertia, the AOUP serves as a minimal reference model to compare and discuss experimental and simulation data.
Second, it allows to understand inertial effects in various environments and settings through obtaining explicit analytical solutions. 
In detail, we give solutions for an inertial AOUP particle affected by constant and harmonic forces and then for two AOUPs connected by a harmonic spring. 
We further explore an active particle which ejects mass in an isotropic way.
A graphical overview of these problems is given in Fig.~\ref{fig_illustration} 
together with an illustration summarizing the different dynamical exponents obtained in this paper.
Parts of our results have been independently obtained recently in Ref.~\cite{inertial}.

This paper is organized as follows.
In Sec.~\ref{sec_model}, we introduce the AOUP model and the dynamical quantities of interest. 
Then we present in Sec.~\ref{sec_results} our main results, elaborating on the role of inertia and the effect of initial conditions,
and conclude in Sec.~\ref{sec_concl}.

\section{inertial AOUP model and noise averages \label{sec_model}}

The active Ornstein-Uhlenbeck particle (AOUP) is arguably the simplest model for one self-propelled particle.
It makes use of a stochastic driving velocity $\mathbf{u}(t)$ with a memory on a finite time scale $\tau$ leading to a persistent motion, which mimics activity.
In detail, this Ornstein-Uhlenbeck process is defined by the stochastic equation
\begin{equation}
\dot{\mathbf{u}}(t) = -\frac{\mathbf{u}(t)}{\tau} + \frac{\bm{\xi}(t)}{\tau}\,, \label{eq_oup}
\end{equation}
where $\bm{\xi}(t)$ is a Gaussian distributed white noise, which is characterized by its first two moments, 
i.e. $\langle \xi_i(t) \rangle = 0$ and $\langle \xi_i(t)\xi_j(t') \rangle = 2D\delta_{ij}\delta(t-t')$ with $i,j \in \{1,...,d\}$ for $d$ spatial dimensions.
Ornstein and Uhlenbeck originally developed the model to study the velocity distribution of passive particles~\cite{uhlenbeck1930theory}, but it can also be used for many other physical and mathematical problems.
Solving Eq.~\eqref{eq_oup} yields the moments for the random velocity $\mathbf{u}(t)$, which is Gaussian distributed colored noise, namely
\begin{align}
\langle u_i(t) \rangle = 0 \quad \text{and} \quad
\langle u_i(t)u_j(t') \rangle = \frac{D}{\tau}\delta_{ij}{\rm e}^{-\frac{|t-t'|}{\tau}}\,. \label{eq_mom}
\end{align}
Here, $\tau$ is the persistence time, which is the time scale at which the stochastic self-propulsion velocity $\mathbf{u}(t)$ randomizes.
The diffusion coefficient $D$ characterizes the motility of the particle.
Both parameters describe the magnitude of the self-propulsion~\cite{AOUPmodel}.
The time scale of the AOUP is $\tau$, so a corresponding length scale can be defined as the persistence length $l_0 \coloneqq \sqrt{D\tau}$.
Finally, the active velocity 
\begin{align}
 u_0:=\sqrt{\langle \mathbf{u}(t)\cdot\mathbf{u}(t) \rangle}=\sqrt{\frac{dD}{\tau}}
\end{align}
can be conveniently related to the equal-time self correlation of $\mathbf{u}(t)$, where $d$ is the spatial dimension.
In the remainder of this work, we restrict ourselves to $d=2$.

The inertial dynamics can be described by the particle's center-of-mass position $\RR(t)$ and velocity $\VV(t)$. 
Given the initial conditions $\RR_0:=\RR(0)$ and $\VV_0:=\dot{\RR}(0)$, we consider the underdamped equation of motion
\begin{equation}
m\ddot{\RR}(t) + \gamma\dot{\RR}(t) = \mathbf{F}_{\text{ext}}(\RR, t) 
+ \gamma\mathbf{u}(t) \label{eq_langevin}
\end{equation}
 for one AOUP in the Langevin picture, where the coefficient of friction for linear drag is denoted by $\gamma$.
Moreover, $\mathbf{F}_{\text{ext}}(\RR, t)=-\vec{\nabla}V_{\text{ext}}(\RR, t)$ 
is an external force caused by an external potential $V_{\text{ext}}(\RR, t)$ acting on the system and $\gamma\mathbf{u}(t)$ represents the active force.
For a fixed activity of the AOUP, the inertial effects can be quantified by defining the dimensionless mass as 
\begin{equation}
 \tilde{m} \coloneqq \frac{m}{\gamma \tau}=\frac{\tau_m}{\tau}\,,
\end{equation}
which can be written as a ratio of two basic time scales, namely the inertial delay time $\tau_m:=m/\gamma$ and the activity persistence time
$\tau$.

As a Gaussian process the AOUP is characterized by its first two moments, Eq.~\eqref{eq_mom}, alone.
To analyze the behavior of such a system, one can calculate dynamical averages and correlations. 
These are the velocity autocorrelation function (VACF) 
\begin{equation}V(t,t'):=\langle \VV(t)\cdot\VV(t') \rangle\,,\label{eq_calcvacf}\end{equation} 
the mean displacement (MD) 
\begin{equation}\mathbf{X}(t):=\langle\RR(t)-\RR_0\rangle\,\end{equation}
and the mean-squared displacement (MSD) 
\begin{equation}\Delta(t):=\langle|\RR(t)-\RR_0|^2\rangle= 2\int_0^t{\rm d}t_1\int_0^{t_1}{\rm d}t_2\,V(t_1,t_2)\,, \label{eq_msdvacf}\end{equation}
where the brackets $\langle\ldots\rangle$ denote a noise average as in Eq.~\eqref{eq_mom}.
 To characterize the dynamical behavior in different time regimes, we introduce the dynamical scaling exponent
\begin{equation}\alpha(t):=\frac{\mathrm{d}\ln(\Delta(t))}{\mathrm{d}\ln(t)}\,,\label{eq_alpha}\end{equation} 
of the MSD. We define the long-time self-diffusion coefficient as
$D_\text{L} \coloneqq \lim_{t \to \infty} \frac{\Delta}{4t}$.
This long-time limit exists in particular, if the dynamical scaling exponent tends to one as $t\to\infty$.

We finally remark that the second moments (MSD and VACF) are the same as for active Brownian particles 
upon identifying $u_0$ with a constant self-propulsion velocity in direction of the instantaneous orientation, 
subject to rotational diffusion with $D_\text{r}=\tau^{-1}$ \cite{farage2015,AOUPmodel} and neglecting translational Brownian diffusion.
Note that, in the AOUP model, a passive Brownian system is conveniently obtained by taking the white-noise limit $\tau\rightarrow0$ of zero persistence time $\tau$ in Eq.~\eqref{eq_oup},
such that the stochastic velocity $\mathbf{u}(t)\equiv\bm{\xi}(t)$ becomes a white noise with the (passive) diffusion coefficient $D$.
For this reason, we do not include an additional white noise in Eq.~\eqref{eq_langevin} to represent the translational Brownian diffusion, usually present in the active Brownian case.

\section{Results \label{sec_results}}

In the following, we determine the solutions of the stochastic differential equation, Eq.~\eqref{eq_langevin} for both $\RR(t)$ and $\VV(t)$ in the scenarios depicted in Fig.~\ref{fig_illustration}.
Then, we calculate different correlation functions by carrying out the noise average with the help of Eq.~\eqref{eq_mom} 
and discuss in detail the time- and mass dependence of the MSD.
To provide the basis for our later study of a harmonic dumbbell and a free particle with linear mass ejection,
we further elaborate on the known results \cite{inertial} for an AOUP in the absence of forces and in a harmonic potential.
Moreover, we consider here a more general nonstationary setup of an AOUP with initial velocity $\VV_0$ and position $\RR_0$ at time $t=0$.
  Selected full analytic solutions of the problems at hand are stated in Appendix \ref{app_calc}.

\subsection{Free particle \label{sec_free}}

As a basic reference, we first consider a free particle in the absence of any external forces $\mathbf{F}_{\text{ext}}=0$. The only relevant time scales
which govern the dynamical correlations are the persistence time $\tau$ and the inertial delay time $\tau_m$.

\subsubsection{Evaluation of analytic solutions \label{sec_free1}}

Solving the equation of motion for the velocity of a free particle,
we find the general VACF as described in appendix \ref{app_calc}.
Taking the steady-state limit, the VACF
\begin{equation}
\!\!\lim_{t' \to \infty} V_\text{f}(t+t',t')
= \frac{2\gamma^2 D}{m^2-\gamma^2\tau^2} \left(\frac{m}{\gamma}{\rm e}^{-\gamma t / m} -\tau{\rm e}^{-t/\tau}\right)\!\!\! \label{eq_freevacf}
\end{equation}
decreases exponentially on the two time scales $\tau_m=m/\gamma$ and $\tau$, independent of the initial velocity $\VV_0$ \cite{inertial}.
The long-time mean-squared velocity
\begin{equation}
\lim_{t \to \infty} V_\text{f}(t,t)
= \frac{2\gamma D}{m + \gamma \tau} = \frac{u_0^2}{\tilde{m}+1} \label{eq_freemsv}
\end{equation}
reflects that heavier particles have on average smaller absolute velocities than lightweight particles which is a clear manifestation of inertia.
The MD 
\begin{equation}
\mathbf{X}_\text{f}(t)= - \frac{m \VV_0}{\gamma} \left( {\rm e}^{-\frac{\gamma t}{m}}-1\right)=\VV_0\,t  + \mathcal{O}(t^2)\,
\label{eq_freemd}
\end{equation} 
 does not depend on the activity since we consider here the stationary active velocity $\mathbf{u}(t)$ with the moments given by Eq.~\eqref{eq_mom}, lacking an initial direction.
 Instead, the MD reflects a persistent motion of particles with a finite initial velocity $\VV_0$ on the inertial time scale, i.e., for $t<\tau_m$.
 For later times, it takes a constant value $\lim_{t \to \infty}\mathbf{X}_\text{f}(t)=\frac{m \VV_0}{\gamma}$ determined by the magnitude and direction of $\VV_0$.
  This finding again constitutes a clear signature of inertia.

 Now we turn to the MSD which we split as
\begin{align}\label{eq_MSDfreeFULL}
\Delta_\text{f}(t)=\Delta^\text{\!\!(ss)}_\text{f}(t)+\Delta^\text{\!\!(acc)}_\text{f}(t)+\Delta^\text{\!\!(0)}_\text{f}(t)\,,
\end{align}
  in terms of the stationary solution 
   \begin{align}\label{eq_MSDfreeFULLss}
\!\!\!\!\!\Delta^\text{\!\!(ss)}_\text{f}=&\,\frac{4D \left[m^3\left({\rm e}^{-\frac{\gamma t}{m}}-1+\frac{\gamma t}{m}\right)-\gamma^3\tau^3\left({\rm e}^{-\frac{t}{\tau}}-1+\frac{t}{\tau}\right) \right]}{\gamma m^2-\gamma^3\tau^2}\!\!\!\!\!\!\!\!\!\!\!\!\!\!\!\cr
=&\,\frac{2 \gamma D}{m + \gamma\tau}\,t^2 - \frac{\gamma^2D}{6m\tau(m + \gamma\tau)}\,t^4 
+ \mathcal{O}(t^5)\!\!\!\!\!\!\!\!\!\!
\end{align}
  for the MSD~\cite{inertial}, a correction term
  \begin{align}\nonumber
\Delta^\text{\!\!(acc)}_\text{f}=&-\frac{2mD\left({\rm e}^{-\frac{\gamma t}{m}}-1\right)}{\gamma m^2-\gamma^3\tau^2} \left[ 
m(m+\gamma\tau)\left({\rm e}^{-\frac{\gamma t}{m}}-1\right)\right.\ \ \ \ \ \ \ \ \ \\\nonumber
&\ \ \ \ \ \ \ \ \ \ \ \ \ \ \ \ \ \ \ \ \ \ \ \ \ \ \ \ \ \ \ \left.-2\gamma^2\tau^2\left({\rm e}^{-\frac{t}{\tau}}-1\right)\right]\ \ \ \ \ \ \ \ \ \\
=&-\frac{2 \gamma D}{m + \gamma\tau}\,t^2 + \frac{(3\gamma\tau+4m)\gamma^2D}{6m^2\tau(m + \gamma\tau)}\,t^4+ \mathcal{O}(t^5)\,, \!\!\!\!\!\!\!\!\!\!\!\!\!\!\!\!
\label{eq_MSDfreeFULLacc}
\end{align}
 initially decreasing the MSD to describe the acceleration of a massive particle starting from rest, and a purely inertial term
\begin{align}\label{eq_MSDfreeFULL0}
\Delta^\text{\!\!(0)}_\text{f} =\mathbf{X}_\text{f}\cdot \mathbf{X}_\text{f}=&\,\frac{m^2\VV_0^2}{\gamma^2}\left({\rm e}^{-\frac{\gamma t}{m}}-1\right)^2\cr
=&\,\VV_0^2\,t^2 - \frac{\gamma \VV_0^2}{m}\,t^3 + \frac{7\gamma^2 \VV_0^2}{12m^2}\,t^4+ \mathcal{O}(t^5)\,,\ \ \
\end{align}
 reflecting the persistence of a general nonzero initial velocity $\VV_0$, just as the MD $\mathbf{X}_\text{f}$ stated in Eq.~\eqref{eq_freemd}.

 The two nonstationary contributions $\Delta^\text{\!\!(acc)}_\text{f}$ and $\Delta^\text{\!\!(0)}_\text{f}$ to the MSD vanish for zero mass and become constant after a long time.
Therefore, both the overdamped limit
 \begin{align}\label{eq_MSDfreeOVERDAMPED}
\lim_{m\to 0} \Delta_\text{f} = 4D \tau\left({\rm e}^{-\frac{t}{\tau}}-1+\frac{t}{\tau}\right)
= \frac{2D}{\tau}\,t^2 - \frac{2D}{3\tau^2}\,t^3 + \mathcal{O}(t^4)
\end{align}
of the MSD and the long-time self-diffusion coefficient $D_\text{L}=D$ follow from $\Delta^\text{\!\!(ss)}_\text{f}$ alone.
Hence, the diffusive behavior of a free inertial AOUP in the long-time limit is mass-independent, as also found for ABPs \cite{scholz2018inertial}.
 Since the quadratic terms in the short-time expansions of $\Delta^\text{\!\!(ss)}_\text{f}$  and $\Delta^\text{\!\!(acc)}_\text{f}$ cancel,
 the early behavior of the MSD is determined by $\Delta^\text{\!\!(0)}_\text{f}$ from Eq.~\eqref{eq_MSDfreeFULL0}.
For an AOUP which is initially at rest, we find 
 \begin{equation}\label{eq_MSDfreeSTrest}
\!\!\!\left.\Delta_\text{f}\right|_{\VV_0=\mathbf{0}}= \frac{\gamma^2D}{2\tau m^2}\,t^4 - \frac {(5\gamma\tau+2m)\gamma^2D}{15\tau^2 m^3}\,t^5 + \mathcal{O}(t^6)\,,
\end{equation}
 which means that it is accelerated on average by $\gamma u_0/m$, where $\gamma^2 u_0^2$ is the average squared activity force.
 The corresponding expansion in the white-noise limit reads
\begin{align}\label{eq_MSDfreeSTpassiverest}
\lim_{\tau \to 0} \left.\Delta_\text{f}\right|_{\VV_0=\mathbf{0}} =& \,\frac{4\gamma^2 D}{3m^2}\, t^3-\frac{\gamma^3D}{m^3}\,t^4 
+ \mathcal{O}(t^5)
\end{align}
  and describes the motion of an initially resting passive particle \cite{breoni2020}.

\begin{figure*} [t] \centering
\includegraphics[width=\textwidth] {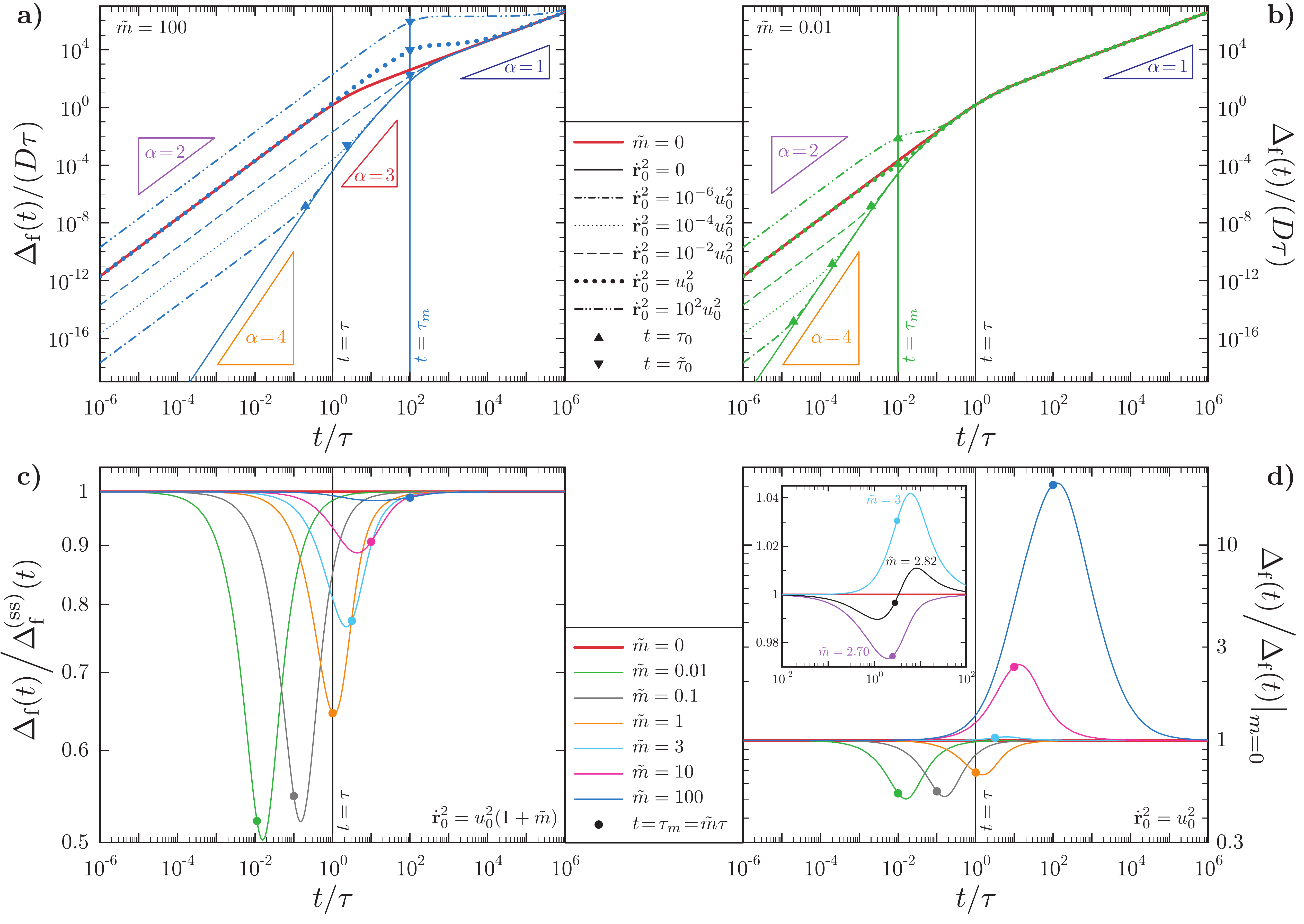}
\caption{ MSD $\Delta_\text{f}(t)$ of a free inertial AOUP, given by Eq.~\eqref{eq_MSDfreeFULL}, with different initial velocities $\VV_0^2$ and masses $\tilde{m}=m/(\gamma \tau)$ (according to labels and legends) compared to the overdamped limit (thick red lines).
The relevant time scales discussed in the text are highlighted (as labeled) where appropriate.
\textbf{a)} MSD for fixed $m=100\gamma\tau$, such that $\tau_m>\tau$. 
\textbf{b)} MSD for fixed $m=0.01\gamma\tau$, such that $\tau_m<\tau$. 
\textbf{c)} MSD for fixed initial velocity $\VV_0^2=u_0^2/(1+\tilde{m})$, chosen to match the stationary mean-squared velocity,
relative to the stationary MSD $\Delta^\text{\!\!(ss)}_\text{f}(t)$ given by Eq.~\eqref{eq_MSDfreeFULLss}.
\textbf{d)} MSD for fixed $\VV_0^2=u_0^2$ relative to that in the overdamped limit, $m\rightarrow0$, given by Eq.~\eqref{eq_MSDfreeOVERDAMPED}.
}
\label{fig_free}
\end{figure*}

\subsubsection{General discussion of the MSD \label{sec_free2}}

The MSD of a free AOUP is graphically evaluated in Fig.~\ref{fig_free} for different parameters.
Comparing both time scales involved, we observe two scenarios.
First, if $\tau_m>\tau$ (or $\tilde{m}>1$, compare Fig.~\ref{fig_free}a), the onset of the long-time diffusive regime with $D_\text{L}=D$ occurs at $t>\tau_m$ and is thus delayed by inertial effects, when compared to the overdamped limit.
Second, if $\tau_m<\tau$ (or $\tilde{m}<1$, compare Fig.~\ref{fig_free}b), there is a ballistic regime due to the persistent active motion
 for $\tau_m<t<\tau$ and the long-time diffusive regime is finally approached for $t>\tau$. 
 More specifically, for $t>\tau_m$, the MSD generally behaves like in the overdamped limit, as given by Eq.~\eqref{eq_MSDfreeOVERDAMPED}.
 
  As also shown in Figs.~\ref{fig_free}a and b, the behavior of the MSD in the early inertial regime for $t<\tau_m$ crucially depends on the ratio between the initial velocity $\VV_0$ and
 the long-time mean-squared velocity of the AOUP, given by Eq.~\eqref{eq_freemsv}, which indicates whether the
 AOUP must (on average) be accelerated or decelerated to reach the stationary state.
  For a sufficiently large $\VV_0^2\gtrsim u_0^2/(1+\tilde{m})$, the whole regime is governed by ballistic motion, according to Eq.~\eqref{eq_MSDfreeFULL0}. 
In the special case $\VV_0^2=u_0^2/(1+\tilde{m})$, the MSD closely follows that in the stationary state, as illustrated in Fig.~\ref{fig_free}c.
The deviations around $t=\tau_m$, become negligible for a large mass.
This can be understood from the short-time expansion in Eq.~\eqref{eq_MSDfreeFULLss},
and the fact that the MSD approaches overdamped behavior after the decay of inertial effects.
If the initial velocity $\VV_0^2\lesssim u_0^2/(1+\tilde{m})$ is even smaller, the initial ballistic regime ends prematurely,
as the AOUP is further accelerated. 

To generally quantify the end of the initial ballistic regime,
we introduce the time scale
\begin{align}
 \tau_0:=\min\left(2\tau_m\frac{|\VV_0|}{u_0},\tau_m\right),
  \label{eq_tau0}
\end{align}
which indicates the onset of an acceleration due to the average activity force
and thus follows from equating the leading terms in the short-time expansions from Eq.~\eqref{eq_MSDfreeSTrest} and Eq.~\eqref{eq_MSDfreeFULL0}, making use of the definition $u_0=\sqrt{2D/\tau}$.
The corresponding superballistic regime with $\alpha=4$ is then observed in both Fig.~\ref{fig_free}a and Fig.~\ref{fig_free}b, 
for $\tau_0<t<\tau<\tau_m$ and $\tau_0<t<\tau_m<\tau$, respectively. 
In the former case, the exponent changes to $\alpha=3$, following Eq.~\eqref{eq_MSDfreeSTpassiverest}, in the regime $\tau_0<\tau<t<\tau_m$,
since the active velocity decorrelates at $t=\tau$.
Moreover, if $\tau_0>\tau$, its role is taken by the alternative time scale
\begin{align}
 \tilde{\tau}_0:=\min\left(\frac{3\tau_m^2}{2\tau}\frac{\VV_0^2}{u_0^2},\tau_m\right),
 \label{eq_ttau0}
\end{align}
deduced from Eq.~\eqref{eq_MSDfreeSTpassiverest} and Eq.~\eqref{eq_MSDfreeFULL0}.
Then, for $\tau<\tilde{\tau}_0<t<\tau_m$, there is a direct transition from the initial ballistic regime to $\alpha=3$,
as visible in Fig.~\ref{fig_free}a.
If $\tau_0=\tau_m$ or $\tilde{\tau}_0=\tau_m$, there is no acceleration regime.

Finally, we consider the special case, $|\VV_0|=u_0$, that the absolute value of the initial velocity equals the active velocity. 
 As highlighted in Fig.~\ref{fig_free}d, the MSD closely resembles the overdamped result for both $t\ll\tau_m$ and $t\gg\tau_m$,
as the quadratic term in the respective short-time expansion from Eq.~\eqref{eq_MSDfreeFULL0} and Eq.~\eqref{eq_MSDfreeOVERDAMPED} is the same. 
The time- and mass-dependent deviation can be inferred from the cubic terms, which become equal for $m=3\gamma\tau$.
For $m>3\gamma\tau$, we observe $\Delta_\text{f}\geq\lim_{m\rightarrow0}\Delta_\text{f}$ for all times,
which merely reflects the implied condition $\tau_m>\tau$, i.e., the ballistic regime due to the persistent initial velocity is longer than
that due to persistent active motion in the overdamped limit, compare Fig.~\ref{fig_free}a.
In contrast, for $m<3\gamma\tau$, the ratio $\Delta_\text{f}/\lim_{m\rightarrow0}\Delta_\text{f}$ first decreases and then returns to unity when the inertial effects have fully relaxed, even if $\tau_m\geq\tau$.
 This behavior indicates that the initial velocity starts to decorrelate at an earlier time than the active motion.
The same can be inferred for the whole duration of both decorrelation processes, regarding in Fig.~\ref{fig_free}d the situation for a mass slightly below $3\gamma\tau$.
In the case $\tau_m<\tau$, where $\Delta_\text{f}\leq\lim_{m\rightarrow0}\Delta_\text{f}$ for all times, we observe in Fig.~\ref{fig_free}b 
 two ballistic regimes, separated at $t=\tau_m$, which both possess the same mean-squared velocity $u_0^2$ but for the two distinct physical reasons discussed before.

\subsubsection{Summary and interpretation of the results \label{sec_free3}}

 Our observations for a free AOUP are summarized in the first column of Fig.~\ref{fig_illustration}.
This schematic exponent diagram should be understood as follows.
The initial regime with $\alpha=2$ is always present (if $\tau_0>0$) and thus belongs to the uppermost layer.
As we have $\tau_0\leq\tau_m$ per definition, these two time scales are drawn on the same layer
Therefore, there are three possibilities for the subsequent dynamical regimes.
First, if $\tau_0<\tau<\tau_m$, as depicted in the illustration, the sequence 2--4--3--1 of exponents $\alpha$
is given by the solid lines.
Second, if $\tau<\tau_0<\tau_m$, which corresponds, e.g., to shifting the vertical bar for $\tau_0$ to the right, 
the regime for $t<\tau$ in the second layer indicating $\alpha=4$ is completely overlaid, such that the sequence is just 2--3--1. 
Third, if $\tau_0<\tau_m<\tau$, which corresponds, e.g., to shifting the vertical bar for $\tau_m$ to the left,
the dotted lines between the old and new position of $\tau_m$ indicate the valid exponent, such that the sequence is 2--4--2--1. 
Further sequences are possible if two or more time scales are equal.
In this qualitative picture $\tau_0$ generally represents the time at which the initial velocity ceases to be persistent.
If one is interested in the explicit formula it should be read as either $\tau_0$ or $\tilde{\tau}_0$, depending on whether $\alpha$ changes to 3 or 4, as discussed in Sec.~\ref{sec_free2}.

 Even in the most simplistic scenario without external forces, the MSD of an AOUP provides deep insights into the fundamental interplay of activity and inertia.
In addition to the results apparent from Fig.~\ref{fig_free}, let us emphasize that the activity enters implicitly through the scaling factors $D$ and $\tau$. 
The effects of increasing the activity thus generally include (i) increasing values for the MSD, (ii) a delay of the onset of the diffusive regime and 
(iii) an effective reduction of the dimensionless mass $\tilde{m}$ (and thus of inertial effects in general),
which should be kept in mind when regarding the following more complex scenarios.

\subsection{Constant force \label{sec_const}}

Next we consider the case of a constant external force ($\mathbf{F}_\text{ext}=\mathbf{F}$ with $F=|\mathbf{F}|$ in Eq.~\eqref{eq_langevin}). 
The steady-state VACF and mean-squared velocity only differ from the free-particle results stated in Eq.~\eqref{eq_freevacf} and~\eqref{eq_freemsv} by the constant term $F^2/\gamma^2$.
The mean displacement $\mathbf{X}_\text{c}(t)$ can be written as  
\begin{equation}
\mathbf{X}_\text{c}-\mathbf{X}_\text{f}=\frac{m \mathbf{F}}{\gamma^2} \left({\rm e}^{-\frac{\gamma t}{m}}-1+\frac{\gamma t}{m}\right)=\frac{\mathbf{F}}{2m}\,t^2  + \mathcal{O}(t^3)\,.
\label{eq_linmd}
\end{equation} 
Hence, $\mathbf{X}_\text{c}$ deviates from the MD of a free particle given in Eq.~\eqref{eq_freemd} by a term 
which denotes an additional acceleration at short times and increases linearly in the long-time limit due to the directed linear force.
  As for a free particle, the pure MD does not carry a footprint of activity under our assumption of a stationary active velocity.

Likewise, the MSD $\Delta_\text{c}(t)$ of an AOUP in a constant force field is supplemented only by terms made up from activity-independent contributions 
that can be expressed in terms of the MD from Eq.~\eqref{eq_freemd} and Eq.~\eqref{eq_linmd}
\begin{align} \label{eq_MSDlinFULL}
\Delta_\text{c}&-\Delta_\text{f}= \mathbf{X}_\text{c}\cdot\left(\mathbf{X}_\text{c}-\mathbf{X}_\text{f}\right)\cr
&= 
\VV_0\cdot\mathbf{F}\left(\frac{1}{2m}\,t^3-\frac{5\gamma}{12m^2}\,t^4\right)+
\frac{F^2}{4m^2}\,t^4  + \mathcal{O}(t^5)\,.\ \ \ \ \ \ 
\end{align} 
 While these additional terms including the constant force $\mathbf{F}$ do not affect the MSD in the ballistic regime with persistent initial velocity $\VV_0$ for $t<\tau_0$, compare Eq.~\eqref{eq_MSDfreeFULL0},
the constant force further enhances the subsequent acceleration due to activity, which shortens the crossover time $\tau_0$ or $\tilde{\tau}_0$ compared to the values given in Eq.~\eqref{eq_tau0} or Eq.~\eqref{eq_ttau0}, respectively, for a free particle.
Moreover, the dynamical exponent in the passive acceleration regime ($\tau<t<\tau_m$)
may change from $\alpha=3$ according to Eq.~\eqref{eq_MSDfreeSTpassiverest}
to $\alpha=4$ when Eq.~\eqref{eq_MSDlinFULL} becomes dominant.
Comparing these expansions, we predict that this happens at $\tau_\text{c}=16\gamma^2D/(3mF^2+12\gamma^3D)$  (if $\tau<\tau_\text{c}<\tau_m$).
The long-time limit $\Delta_\text{c}\simeq (F^2/\gamma^2)t^2$ of the MSD is always ballistic with velocity $F/\gamma$.
This final regime surpasses a free-particle-like diffusive regime with $\Delta_\text{c}\simeq 4Dt$ for $t>\tilde{\tau}_\text{c}=4D\gamma^2/F^2$ if $\tilde{\tau}_\text{c}>\tau$ and $\tilde{\tau}_\text{c}>\tau_m$. 

All possible dynamical exponents are illustrated in the second column of Fig.~\ref{fig_illustration},
where $\tau_\text{c}$ should be read as $\tilde{\tau}_\text{c}$ if the inertial time scale $\tau_m$ becomes shorter, as described above.
We also see that in the case $\tau_0<\tau_m<\tau<\tilde{\tau}_\text{c}$ there are three distinct ballistic regimes due to persistent inertial motion with initial velocity $\VV_0$, persistent active motion and, finally, the constant external force.

\begin{figure*} [t] \centering
\includegraphics[width=\textwidth] {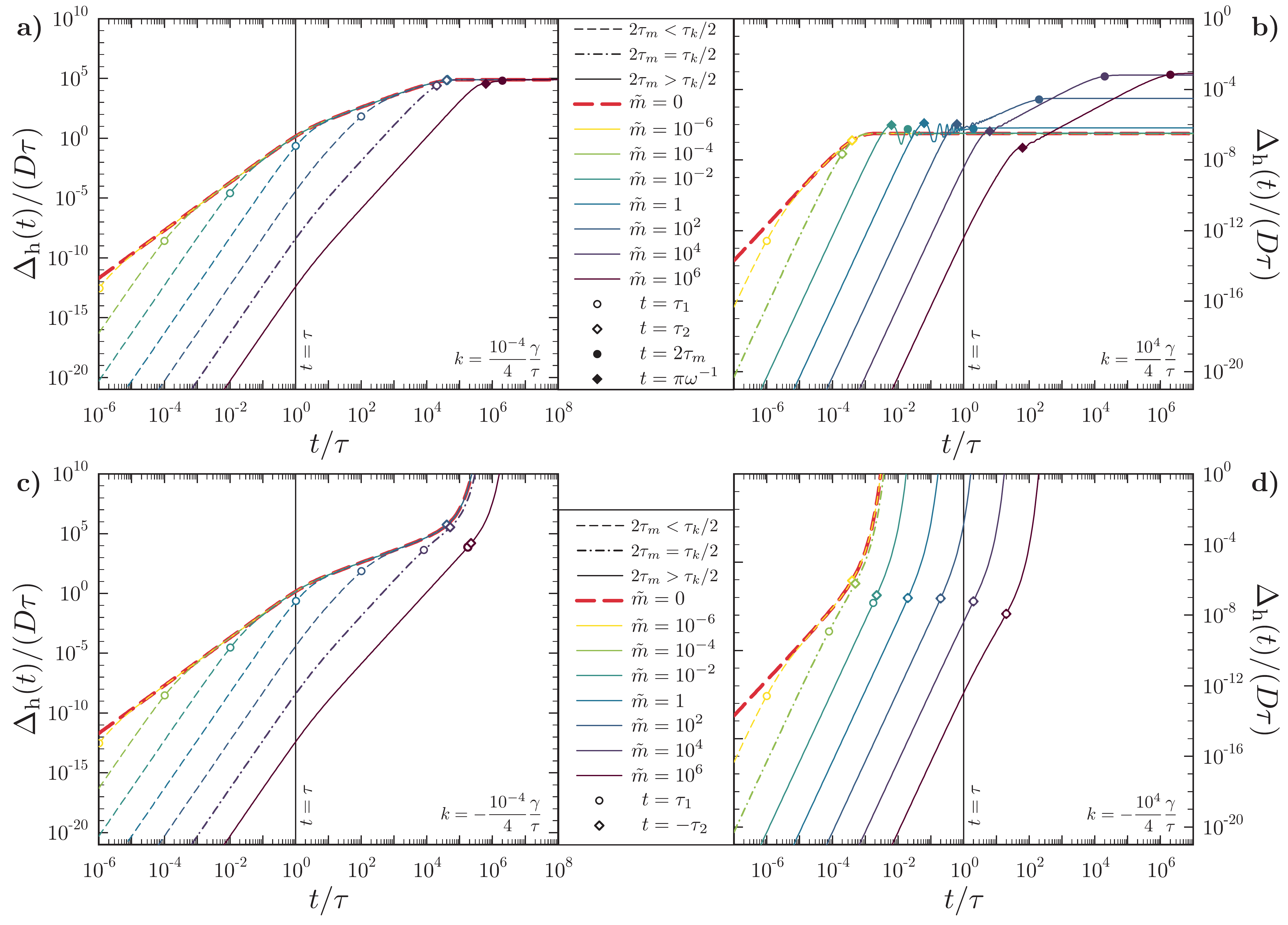}
\caption{ MSD of an inertial AOUP initially resting ($\VV_0=\mathbf{0}$) in the center ($\RR_0=\mathbf{0}$) of a harmonic potential, Eq.~\eqref{eq_harmonicpot}, with constant $k$. 
We consider different masses $\tilde{m}=m/(\gamma \tau)$ (according to legends) and compare to the overdamped limit (thick red lines).
 \textbf{a)} MSD in a trap with $k=10^{-4}\gamma/(4\tau)$, such that $\tau_k>\tau$ and critical damping (dot-dashed lines) for $m=10^4\gamma\tau$.
\textbf{b)} MSD in a trap with $k=10^{4}\gamma/(4\tau)$, such that $\tau_k<\tau$ and critical damping (dot-dashed lines) for $m=10^{-4}\gamma\tau$.
\textbf{c)} MSD in an unstable potential with $k=-10^{-4}\gamma/(4\tau)$.
\textbf{d)} MSD in an unstable potential with $k=-10^{4}\gamma/(4\tau)$.
}
\label{fig_harm}
\end{figure*}

 \subsection{Harmonic potential \label{sec_harm}}
 
 As a next step we consider an AOUP subject to a time-independent external force in Eq.~\eqref{eq_langevin} generated by the harmonic potential 
 \begin{align}
 V_{\text{ext}}(\RR)=\frac{1}{2}k\RR^2 
 \label{eq_harmonicpot}
 \end{align}
 with the constant $k$. We consider here both cases of a harmonic trap, where $k>0$ acts as a spring constant,
 and an unstable situation with $k<0$. 
   For such a nonlinear potential the translational invariance is broken,
   such that the noise-averaged quantities of interest 
   explicitly depend on the initial position $\RR_0$.

   Here we focus on the MSD $\Delta_\text{h}$, for which we obtain the general short-time expansion 
   \begin{align}
\Delta_\text{h}(t) =&\, \VV_0^2\,t^2 - \frac{\VV_0 \cdot (\gamma\VV_0 + k\RR_0)}{m}\,t^3 \cr
&+ \frac{(7\gamma^2\VV_0^2- 4km\VV_0^2 + 10\gamma k\RR_0 \cdot \VV_0 + 3k^2\RR_0^2 )}{12m^2}\,t^4 \cr
&+ \frac{\gamma^2D}{2m^2\tau}\,t^4 + \mathcal{O}(t^5)\,,
\end{align}
whose leading terms with and without an initial velocity are the same as for a free particle, cf.\ Eq.~\eqref{eq_MSDfreeFULL0} and Eq.~\eqref{eq_MSDfreeSTrest}, respectively.
For vanishing initial conditions $\RR_0=\mathbf{0}$ and $\VV_0=\mathbf{0}$,  the first correction 
\begin{align}
 \Delta_\text{h}(t)-\Delta_\text{f}(t)=-k\frac{\gamma^2D}{12m^3\tau}\,t^6 + \mathcal{O}(t^7)
\end{align}
to the free-particle expansion depending on $k$ appears at sixth order in time.
The sign of this term indicates that the MSD compared to a free AOUP is reduced for a positive $k$, i.e., if the particle starts in the center of a harmonic trap,
and enhanced for a negative $k$, i.e., if the particle initially sits on top of an unstable potential hill.

In the long-time limit, the MSD diverges exponentially for $k<0$, while we find for $k>0$ the expression 
\begin{align}
\lim_{t \to \infty} \Delta_\text{h}(t) = \RR_0^2 + \frac{2(m + \gamma\tau)\gamma D}{k(m + \gamma\tau + k\tau^2)}\,,
\label{eq_hoconvergeud}
\end{align}
which is constant in time and reflects how far (on average) the particle can climb the potential gradient of the trap.
This distance thus increases (i) for an increasing average active velocity $u_0=\sqrt{2D/\tau}$,
(ii) for an increasing persistence of the particle's velocity due to inertia (increasing mass $m$) or activity (increasing persistence time $\tau$ at constant $u_0$)
and (iii) for a decreasing spring constant $k$.
 The initial position of the particle in the potential merely marks a vertical offset.

\begin{figure*} [t] \centering\hspace*{-0.1cm}
\includegraphics[width=\textwidth] {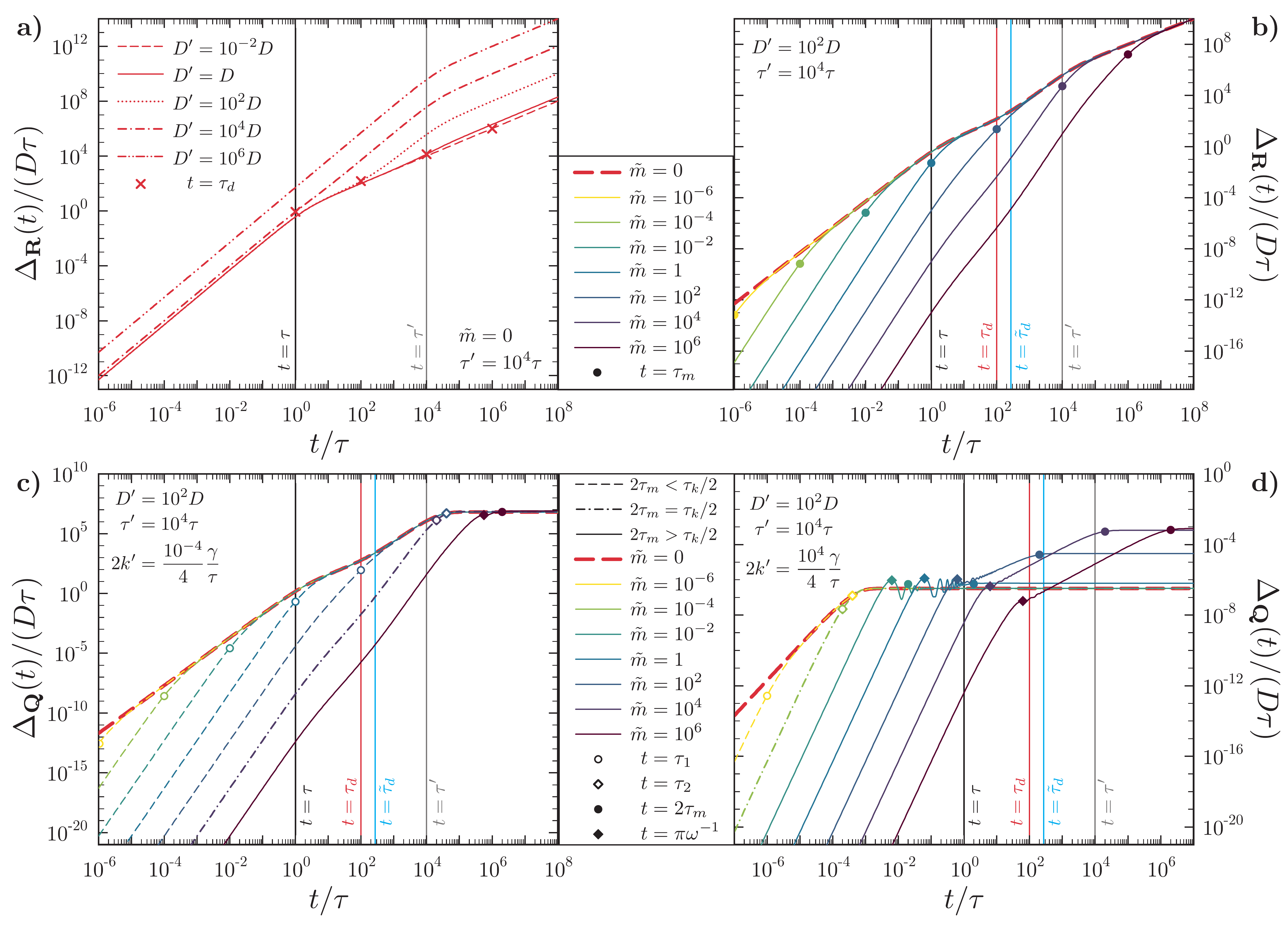}
\caption{ 
\textbf{a,b)} MSD $\Delta_\mathbf{R}$ of the center-of-mass coordinate $\mathbf{R}$
and \textbf{c,d)} MSD $\Delta_\mathbf{Q}$ of the relative coordinate $\mathbf{Q}$ 
for an inertial harmonic dumbbell consisting of two AOUPs, coupled by the spring constant $k'$, with identical masses $\tilde{m}=m/(\gamma \tau)$
and zero initial velocity ($\dot{\mathbf{R}}_0=\dot{\mathbf{Q}}_0=\mathbf{0}$). The second particle may have a different persistence time $\tau'$ and diffusivity $D'$,
which introduces two additional active time scales $\tau'$ and $\tau_\text{d}$ (or $\tilde{\tau}_\text{d}$, see text for details), as indicated by the vertical lines.
All parameter values are labeled or given in the legends.
}
\label{fig_dumb}
\end{figure*}

To understand the full analytic solution for the MSD, given in appendix~\ref{app_calc} and illustrated in Fig.~\ref{fig_harm}, 
we first notice that in the overdamped limit the trap merely induces an additional time scale $\tau_k:=\gamma/|k|$, which indicates how long the particle can (on average) move freely before being affected by the potential.
 For a finite particle mass, the relevant passive time scales
 can be determined from the exponential solutions $\RR(t)\propto\exp(-t/\tau_{1/2})$ of homogeneous differential equation $m\ddot{\RR}(t) + \gamma\dot{\RR}(t) +k\RR(t) = \mathbf{0}$,
 while the active time scale $\tau$ enters through the inhomogeneous part of Eq.~\eqref{eq_langevin}.
 In general, we find
    \begin{align}
\tau_{1/2}
=\frac{2\tau_m}{1\pm\sqrt{1-4\frac{\tau_m}{\tau_k}\,\text{sgn}(k)}}\,,
 \label{eq_roots}
\end{align} 
 where $\text{sgn}(k)$ denotes the sign of $k$.
 Expanding these factors for $\tau_m\ll\tau_k$ yields $\tau_1\simeq\tau_m$ and $\tau_2\simeq\text{sgn}(k)\tau_k$,
 which means that they denote the decay of inertial effects and the onset of potential effects, respectively.
  The detailed behavior depends on the sign of $k$ and is discussed in the following.

The MSD in a harmonic trap with $\text{sgn}(k)=1$ is illustrated in Figs.~\ref{fig_harm}a and b.
 It becomes apparent that the different dynamical regimes are separated by the time scales $\tau$ and $\tau_{1/2}$ from Eq.~\eqref{eq_roots} 
 as long as the particle's mass is below a critical value, determined by the condition $\tau_k/2=2\tau_m$, such that $\tau_1=\tau_2$.
 As long as $t\lesssim\tau_2$, the MSD resembles that discussed in Sec.~\ref{sec_free} for a free particle, which is best observed in Fig.~\ref{fig_harm}a. 
  Unlike the free-particle case, however, the MSD does not revert to the overdamped limit for $t\gg\tau_1$ 
 but rather takes a constant value for long times, which explicitly depends on mass, activity and the spring constant, according to Eq.~\eqref{eq_hoconvergeud}.
 For critical damping, the acceleration regime is directly followed by the final regime with a constant MSD.   
 For even larger masses, we rewrite Eq.~\eqref{eq_roots} as $\tau_{1/2}^{-1}=(2\tau_m)^{-1}\pm i\omega$, introducing the angular frequency
\begin{align}
 \omega := \frac{1}{2 m}\sqrt{4 k m - \gamma^2}= \frac{1}{2 \tau_m}\sqrt{4\frac{\tau_m}{\tau_k} - 1}
 \label{eq_omega}
\end{align}
of the oscillation, such that the MSD for $\VV_0=\mathbf{0}$ develops a first maximum after a half period $\pi\omega^{-1}$,  compare Fig.~\ref{fig_harm}b.
 The time scale $2\tau_m>\omega^{-1}$ then marks the end of the oscillatory regime in this underdamped case, as the inertial persistence ceases and the MSD remains constant.

 The dynamical exponents of an overdamped AOUP in a harmonic trap are summarized in the third column of Fig.~\ref{fig_illustration}, where the indicated time scales represent the overdamped situation.
 In the underdamped case, where the time scale labeled as $\tau_1$ is larger than that labeled $\tau_2$, these labels should be interpreted as $\pi\omega^{-1}$ and $2\tau_m$, respectively.
  Further note that the active time scale $\tau$ does not indicate a change of the dynamical exponent if it is the longest time scale in the system
  but still affects the maximal MSD, given by Eq.~\eqref{eq_hoconvergeud}, in the constant regime.
In the most general scenario with $0<|\VV_0|<u_0$, $k>0$ and $\tau<\tau_m<\tau_k/4$, there are five different dynamical exponents $\alpha\in\{0,1,2,3,4\}$.

  In the case of an unstable potential with $\text{sgn}(k)=-1$, 
 there are always the two exponential time scales $\tau_1$ and $-\tau_2>\tau_1>0$ from Eq.~\eqref{eq_roots}.
 As shown in Figs.~\ref{fig_harm}c and d, the MSD behaves like in the force-free case or in a harmonic trap
  until the particle begins to feel the potential at $t\approx-\tau_2$, which results in the onset of exponential growth.
In contrast to the harmonic trap, the unstable potential has no critical damping.
The equality of $\tau_k/2=2\tau_m$ rather indicates a crossover between the two limits
 $\tau_m\ll\tau_k$, where $-\tau_2\simeq\tau_k$ is mass-independent (and equal to the overdamped limit),
 and $\tau_m\gg\tau_k$, where $-\tau_2\simeq\tau_1\simeq\sqrt{\tau_m\tau_k}$ increases with increasing mass and approaches $\tau_1$.

 \subsection{Harmonic dumbbell \label{sec_dumb}}
 
 As a next step, we consider a generalization of Eq.~\eqref{eq_langevin}
 by introducing another AOUP with the same mass $m$ and an active velocity $\mathbf{u}'(t)$ with the distinct parameters $D'$ and $\tau'$,
 which is coupled to the first particle by a harmonic force of spring constant $k'>0$.
 The coupled Langevin equations describing this setup read 
 \begin{align}
&m \ddot{\RR}_1 + \gamma \dot{\RR}_1 + k' (\RR_1-\RR_2) = \gamma\mathbf{u}(t) \,,\label{eqn:dumb1}\\
&m \ddot{\RR}_2 + \gamma \dot{\RR}_2 + k' (\RR_2-\RR_1) = \gamma\mathbf{u}'(t)\,. \label{eqn:dumb2}
\end{align}
%
 To decouple 
 we transform the coordinates by defining the position of the center of mass as 
 $\mathbf{R}(t) \coloneqq \frac{1}{2}(\RR_1 + \RR_2)$ 
and the relative position of the two particles as $\mathbf{Q}(t) \coloneqq \RR_1 - \RR_2$. 
In these newly defined coordinates Eqs.~\eqref{eqn:dumb1} and~\eqref{eqn:dumb2} become
\begin{align}
m \ddot{\mathbf{R}} + \gamma \dot{\mathbf{R}} &= \frac{1}{2}(\mathbf{f}_1(t)+\mathbf{f}_2(t))\,, \label{eq_dumbR}\\
m \ddot{\mathbf{Q}} + \gamma \dot{\mathbf{Q}} + 2k' \mathbf{Q} &= \mathbf{f}_1(t)-\mathbf{f}_2(t)\,, \label{eq_dumbQ}
\end{align}
with the initial conditions ${\mathbf{R}}_0 = {\mathbf{R}}(0)$, $\dot{\mathbf{R}}_0 = \dot{\mathbf{R}}(0)$, $\mathbf{Q}_0 = \mathbf{Q}(0)$
and $\dot{\mathbf{Q}}_0 = \dot{\mathbf{Q}}(0)$.
In the following, we assume $\tau'\geq\tau$ without loss of generality.

Focusing first on
Eq.~\eqref{eq_dumbR}, we immediately see that the center of mass $\mathbf{R}$ behaves like a free particle subject to two independent random forces.
The corresponding MSD can thus be constructed as 
\begin{align}
 \left.\Delta_\mathbf{R}=\frac{\Delta_\text{f}+\Delta_\text{f}'}{4}\right|_{\VV_0=2\dot{\mathbf{R}}_0}\,,
\end{align}
where $\Delta_\text{f}$ and $\Delta_\text{f}'$ are both given by Eq.~\eqref{eq_MSDfreeFULL} for the respective activity parameters of the two particles.
The center-of-mass motion is subject to the additional time scales $\tau_\text{d}:=D\tau'/D'$ and $\tau'$,
which is best understood in the overdamped limit. 
In this case, Fig.~\ref{fig_dumb}a illustrates that the initial ballistic motion for 
$t<\tau$, determined by the expansion $\Delta_\mathbf{R}=(D/\tau+D'/\tau')t^2+\mathcal{O}(t^3)$, depends on the activity parameters of both particles.
Likewise, we find $\Delta_\mathbf{R}\simeq(D+D')t$ for $t>\tau'$,
which means that the value of the long-time diffusion coefficient $D_\text{L}=(D+D')/4$ of the dumbbell equals half the average of that of two free particles.
The MSD in the intermediate time regime, $\tau<t<\tau'$, is subject to the competition between
the diffusive behavior with $\Delta_\mathbf{R}\simeq Dt$ of the less persistent particle
and the ballistic behavior with $\Delta_\mathbf{R}\simeq (D'/\tau')t^2$ of the more persistent particle.
Equating the two expressions shows that a transition from the former to the latter can be observed at $t=\tau_\text{d}$ if $\tau<\tau_\text{d}<\tau'$.
Otherwise, there are in total only three time regimes, while in the two extreme cases $D'\ll D$ and $D'\gg D$ only the transition from ballistic to diffusive is observable at $t=\tau$ and $t=\tau'$, respectively.

With inertia, the short-time behavior of the MSD differs from the overdamped limit for $t<\tau_m$, in analogy to a free particle.
If the center of mass is initially at rest ($\dot{\mathbf{R}}_0=\mathbf{0}$), Fig.~\ref{fig_dumb}b illustrates
 up to three different superballistic acceleration regimes in the case $\tau<\tilde{\tau}_\text{d}<\tau_m$,
 where the time scale $\tilde{\tau}_\text{d}:=8D\tau'/(3D')$ for a possible transition from the dynamical exponent three to four 
can be found by equating the leading terms in Eq.~\eqref{eq_MSDfreeSTpassiverest} and Eq.~\eqref{eq_MSDfreeSTrest} for the appropriate parameters.
As $\tilde{\tau}_\text{d}\simeq{\tau}_\text{d}$, we observe in general analogy to the MSD of a free particle that the exponents three or four occur for $t<\tau_m$ if the overdamped behavior is diffusive or ballistic, respectively. 
All possible dynamical exponents are illustrated in the fourth column of Fig.~\ref{fig_illustration},
where $\tilde{\tau}_\text{d}$ takes the role $\tau_\text{d}$ if the inertial time scale $\tau_m$ becomes shorter, as described above.

  \begin{figure*} [t] \centering
\includegraphics[width=\textwidth] {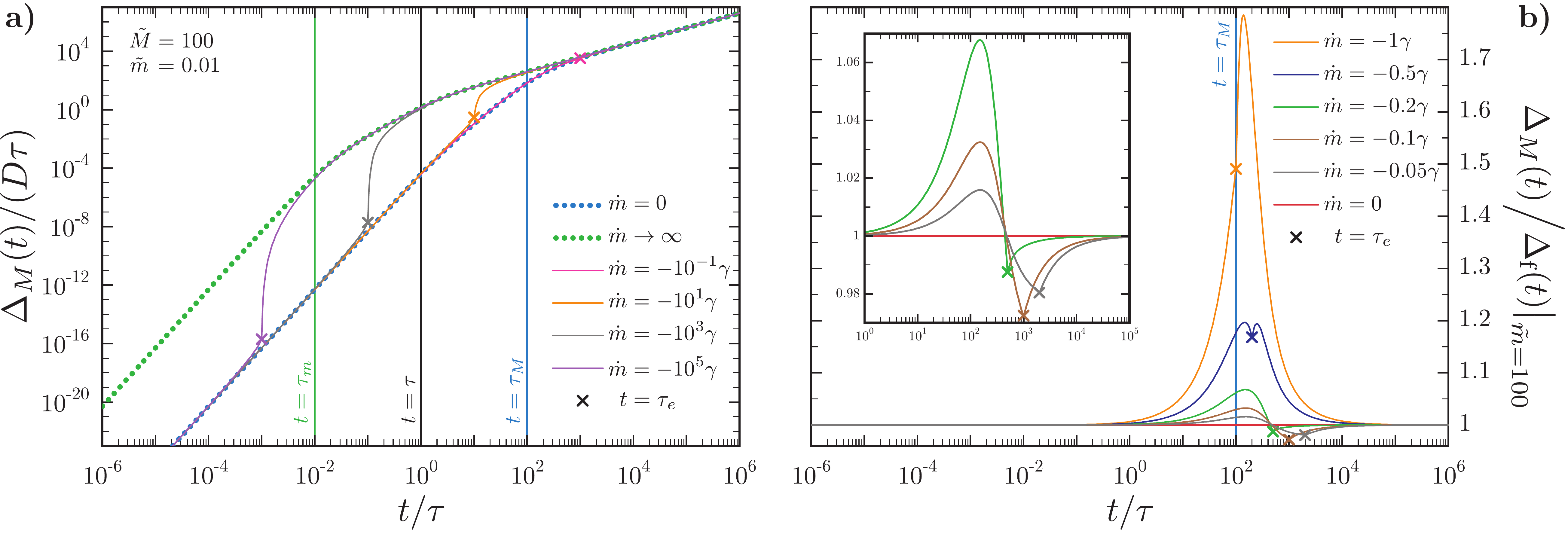}
\caption{
MSD of an inertial AOUP with vanishing initial velocity $\VV_0=\mathbf{0}$ and linear mass ejection $m(t)$, described by Eq.~\eqref{eq_mt}, from initial mass $M=100\gamma\tau$ to final mass $m=0.01\gamma\tau$
with different slopes $\dot{m}<0$ as labeled. The time $\tau_\text{e}$ at which the mass ejection ends is highlighted by crosses.
\textbf{a)} Comparison to the MSD of a free particle with constant mass $m$ or $M$.
\textbf{b)} Relative MSD to that of a free particle with mass $M$ for $\tau_\text{e}\simeq\tau_M$. 
}\label{fig_mass}
\end{figure*}

The relative position $\mathbf{Q}$ of the two monomers evolves in time according to
Eq.~\eqref{eq_dumbQ}, i.e., like a single particle in an effective harmonic trap with spring constant $k=2k'$, compare Eq.~\eqref{eq_harmonicpot}. 
The resulting MSD can thus be expressed as
\begin{align}
 \left.\Delta_\mathbf{Q}=\left(\Delta_\text{h}+\Delta_\text{h}'\right)\right|_{\VV_0=\dot{\mathbf{Q}}_0/2}\,,
\end{align}
where the full expression for $\Delta_\text{h}$ is given in appendix~\ref{app_calc}.
 The relevant time scales $\tau_1'$ and $\tau_2'$, denoting the end of the inertial regime and the onset of confinement effects, respectively, can be inferred from Eq.~\eqref{eq_roots}.
Recalling the discussion from Sec.~\ref{sec_harm}, the second active time scale $\tau'$ is only relevant if it is not the largest time scale, 
which requires a relatively small (effective) coupling between the particles, as illustrated in Fig.~\ref{fig_dumb}c.
In this case, the short-time behavior is similar (up to a factor four) to that of the unbounded center of mass with two active time scales, as discussed in the previous paragraph.
The MSD then becomes constant for times exceeding the threshold which is set by the spring constant or the particles' mass.
The maximal displacement of the relative coordinate can be easily deduced from Eq.~\eqref{eq_hoconvergeud} and depends on all four activity coordinates and the particles' mass.
For a stronger coupling between the particles, Fig.~\ref{fig_dumb}d depicts characteristic oscillations in the relative MSD, whose angular frequency $\omega'$ follows from inserting $k=2k'$ into Eq.~\eqref{eq_roots}.
 In conclusion, there are up to seven different dynamical regimes possible for the relative position of the AOUPs connected to a harmonic dumbbell, covering all dynamical exponents $\alpha$ ranging from zero to four.
This behavior can be illustrated by combining the third and fourth column of Fig.~\ref{fig_illustration}.

 \subsection{Time-dependent mass  \label{sec_mass}}
 
 Our final setup consists of a particle with a time-dependent mass $m(t)$.
 We consider here only an isotropic (undirected) ejection or accumulation of mass, in contrast to the rocket-like setup discussed in Ref.~\cite{rocket}.
 Hence, we start from the generic Langevin equation, Eq.~\eqref{eq_langevin}, by replacing $m$ with $m(t)$ for a free particle with $\mathbf{F}_{\text{ext}}=0$. 
 In particular, to allow for an analytic solution \footnote{The problem of an AOUP with a time-dependent mass $m(t)$ as given by Eq.~\eqref{eq_mt} admits an analytic solution for the MSD in terms of hypergeometric functions, 
 which is too lengthy to be stated here but is available from the authors upon request.},
 we consider the mass to change linearly in time according to the function
  \begin{equation}
m(t)=
\begin{cases}
 M + \dot{m}t\,,\quad & \text{for } t<\frac{m-M}{\dot{m}}
\,,\\
m\,,\quad & \text{for } t\geq\frac{m-M}{\dot{m}}
\,,
\end{cases}\label{eq_mt}
\end{equation}
 where $M \coloneqq m(0)$ is the initial, $m$ the final mass of the particle and $\dot{m}$ denotes the constant time derivative of $m(t)$ in the time-dependent regime.
 The limits $\dot{m}\rightarrow0$ and $\dot{m}\rightarrow\pm\infty$ correspond to a free AOUP with constant mass $M$ and $m$, respectively.
 Moreover, $\dot{m}<0$ denotes the rate of mass ejection and $\dot{m}>0$ the rate of mass accumulation.
 In the remainder of this section, we discuss the MSD $\Delta_M$ of an AOUP for these two cases separately.

  \begin{figure*} [t] \centering 
\includegraphics[width=\textwidth] {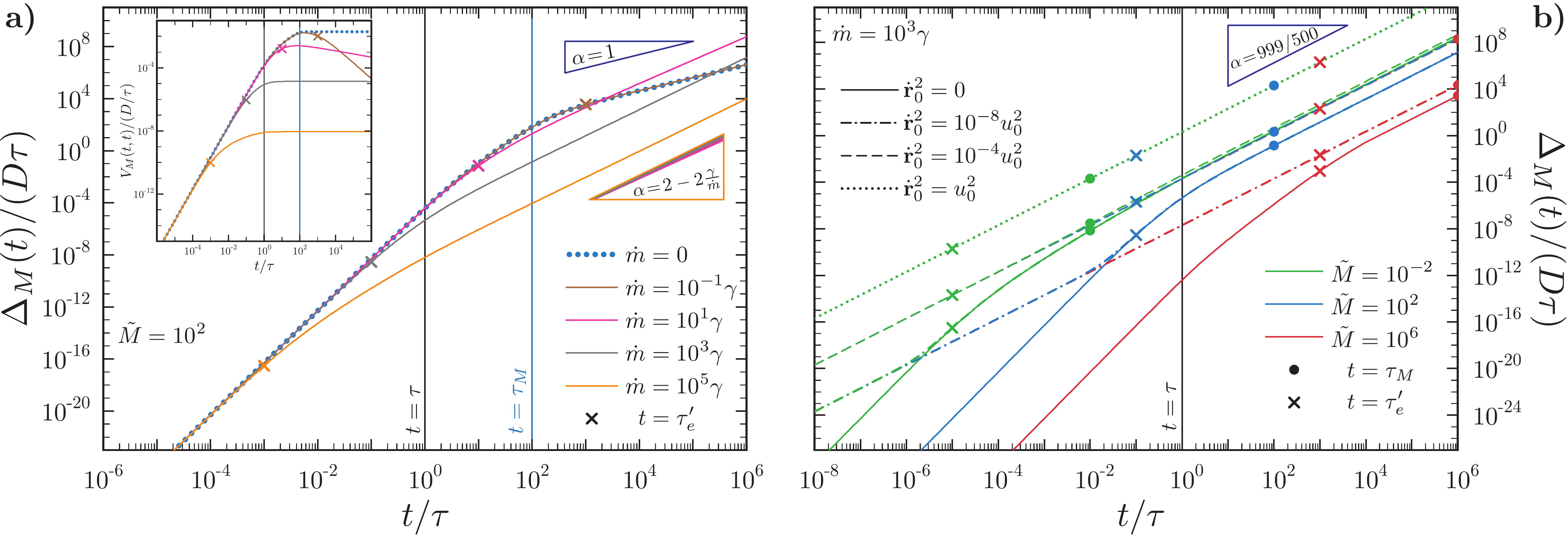}
\caption{
MSD of an inertial AOUP with linear mass accumulation $m(t)$ described by Eq.~\eqref{eq_mt} with different slopes $\dot{m}>0$ as labeled. 
The time $\tau_\text{e}'$ at which the gained mass $m(\tau_\text{e}')-M=2M$ equals twice the initial mass $M$ is highlighted by crosses.
\textbf{a)} Comparison for fixed initial mass $M=100\gamma\tau$ and vanishing initial velocity $\VV_0=\mathbf{0}$. The case $\dot{m}=0$ with constant mass $M$ corresponds to a free particle as in Fig.~\ref{fig_free}a.
\textbf{b)} Comparison of the MSD with constant accumulation rate $\dot{m}=10^3\gamma$ for different initial parameters $M$ and $\VV_0^2$ as labeled.
 The lines corresponding to the same $\VV_0^2$ lie partially on top of each other.
}\label{fig_mass2}
\end{figure*}

 \subsubsection{Mass ejection}

 An AOUP whose mass decreases linearly in time according to Eq.~\eqref{eq_mt} 
 is affected by this process until all ''fuel'' of mass $M-m$ is depleted at the characteristic time $\tau_\text{e}:=\frac{m-M}{\dot{m}}$.
Afterwards it behaves as a free particle of mass $m$. 
  Hence, for long times, the MSD generally reverts to the overdamped result.
 In terms of maximizing the MSD, the strategy to eject fuel gives a temporary advantage compared to moving with constant initial mass $M$ 
if the initial velocity $|\VV_0|\ll u_0$ is so small that the AOUP first needs to be accelerated, which we illustrate for an initially resting particle ($\VV_0=\mathbf{0}$) in
Fig.~\ref{fig_mass}a.
The particular relevance of the possible inertial timescales $\tau_M=M/\gamma$ or $\tau_m=m/\gamma$ of the particle with or without fuel
is therefore closely related to the 
time scale $\tau_\text{e}$
at which the change of mass takes effect.

In detail, for $\tau_\text{e}\gtrsim\tau_M$, i.e., a slow mass ejection, the overall MSD is largely the same as for a free AOUP with constant initial mass $M$, as the behavior for $t>\tau_M$ is not strongly affected by the particle's mass.
 The behavior in this time regime is emphasized in Fig.~\ref{fig_mass}b, which also illustrates the initially enhanced acceleration due to mass ejection.
Moreover, the MSD for a sufficiently slow mass ejection eventually falls below the
MSD for constant $M$, with a maximal relative deviation at $t=\tau_\text{e}$,
which is because the velocity decorrelates earlier for a smaller mass $m(t)<M$, before the common overdamped limit is approached.
For a faster mass ejection, the time scale $\tau_\text{e}$ indicates an exponential approach of the MSD to that of a free particle with mass $m$,
as apparent from the nearly vertical lines in Fig.~\ref{fig_mass}a.
Hence, for $\tau_m\lesssim\tau_\text{e}\lesssim\tau_M$, the inertial regime ends abruptly at $t=\tau_\text{e}$, as the MSD directly switches from underdamped behavior with mass $M$ to the overdamped result.
Finally, for $\tau_\text{e}\lesssim\tau_m$, there is a transition at $t=\tau_\text{e}$ between two different superballistic regimes with the same dynamical exponents, as the magnitude of the average acceleration decreases due to the lost mass.
The inertial regime then ends at $t=\tau_m$.

The dynamical exponents for an AOUP with linear mass ejection are illustrated in the fifth column of Fig.~\ref{fig_illustration},
 where the dotted vertical bar labeled $\tau_\text{e}$ is only relevant for $\tau_\text{e}<\tau_m$.
If the time $\tau_\text{e}>\tau_m$ of mass ejection takes longer than the inertial time scale of the empty particle,
then the label $\tau_m$ should be read as $\min\left(\tau_\text{e},\tau_M\right)$.
 This general exponent diagram also emphasizes that there is no effect of mass ejection observable (compared to a free particle with empty mass $m$) 
if the particle starts with a finite initial velocity $\VV_0$ such that $\tau_0>\tau_\text{e}$.
This can be understood from the short-time expansion 
\begin{align}\label{eq_MSDfreeFULL0e} 
\Delta^\text{\!\!(0)}_M =\,\VV_0^2\,t^2 - \frac{\gamma \VV_0^2}{M}\,t^3 + \frac{(7\gamma+4\dot{m})\gamma \VV_0^2}{12M^2}\,t^4+ \mathcal{O}(t^5)\,,
\end{align}
of the term depending on $\VV_0$, generalizing Eq.~\eqref{eq_MSDfreeFULL0},
whose leading order does not depend on the mass.
 Finally, we stress that for a large initial velocity $|\VV_0|\gg u_0$ 
the strategy of mass ejection results in a general disadvantage compared to moving with constant initial mass $M$,
since the direction of $\VV_0$ remains persistent for a shorter time
if the mass is depleted.

\subsubsection{Mass accumulation}

For an AOUP whose mass increases linearly over time according to Eq.~\eqref{eq_mt},
we focus on the particular limit $m\rightarrow\infty$ that the mass accumulation continues indefinitely.
Next, we introduce the timescale $\tau_\text{e}':=\frac{2M}{\dot{m}}$ denoting the time when the particle has accumulated the double amount of its initial mass $M$
and examine its competition with the second inertial time scale $\tau_M=M/\gamma$.
The typical behavior of the MSD is shown in Fig.~\ref{fig_mass2}a for an initially resting particle.
The situation for a finite value of $m$ can be easily inferred by appreciating
that the behavior reverts to the generic overdamped limit not later than $t=m/\gamma$, in analogy to earlier discussions.
 For $m\rightarrow\infty$, however, the MSD does not necessarily revert to overdamped behavior, as we discuss below.

In analogy to the ejection case, the MSD is qualitatively similar to that of a free AOUP with constant mass $M$ if $\tau_\text{e}'\gtrsim\tau_M$, 
which means that the accumulation of mass happens not fast enough to delay or even prevent the end of the inertial regime.
 Thereafter, the particle's motion does not become stationary, as its mean-squared velocity 
\begin{equation}
V_M(t,t)\stackrel{t\gg\tau_M}{=}
 \frac{2\gamma D}{m(t) + \gamma \tau}\ \ \ \ (\mbox{if}\ \tau_\text{e}'>\tau_M) \label{eq_freemsvME}
\end{equation}
 continuously decreases for long times, adiabatically following the free-particle result from Eq.~\eqref{eq_freemsv}.
In the opposite case, for $\tau_\text{e}'<\tau_M$, Fig.~\ref{fig_mass2}a shows that the slope of the MSD decreases as the particle becomes increasingly massive for $t\gtrsim\tau_\text{e}'$,
reflecting its retarded acceleration.
   The maximal velocity, once reached, then remains nearly persistent, as the acceleration due to random forces, 
 which aim to disperse the particle's direction of motion, becomes more and more irrelevant with increasing mass.
 This is best reflected in the particle's mean-squared velocity $V_m(t,t)$,  shown in the inset of Fig.~\ref{fig_mass2}a, for large rates $\dot{m}$ of mass accumulation.
This balance eventually leads to superdiffusive but subballistic behavior in the long-time limit, i.e., the inertial regime never ends if $\tau_\text{e}'<\tau_M$.
The corresponding dynamical exponent 
\begin{align}
 \alpha=2-\frac{\tau_\text{e}'}{\tau_M}\ \ \ \ (\mbox{if}\ \tau_\text{e}'<\tau_M)\,,
 \label{eq_alphaaccu}
\end{align} 
can be determined analytically in the white-noise limit (which is generally recovered for $t\gg\tau$)
and is numerically confirmed for all curves shown in Fig.~\ref{fig_mass2}.
Therefore, the MSD for strong mass accumulation eventually  surpasses  that of an AOUP with $\tau_\text{e}'>\tau_M$, which becomes diffusive ($\alpha=1$) at $t=\tau$ or $t=\tau_M$.
 In the special case $\tau_\text{e}'=\tau_M$, the MSD behaves as $\Delta_M\simeq t\ln(t)$ for long times.

Apart from the modified dynamical exponent in Eq.~\eqref{eq_alphaaccu},
the long-time behavior in the case $\tau_\text{e}'<\tau_M$  depends on both initial mass $M$ and velocity $\VV_0$, as illustrated in Fig.~\ref{fig_mass2}b, and (implicitly) also on the active velocity $u_0=\sqrt{2D/\tau}$.
This observation is related to the particle's maximal (persistent) velocity, which follows from these parameters.
 Therefore, the MSD at long times is generally enhanced for smaller $M$ and higher $u_0$, which both increase the initial acceleration as long as $|\VV_0|\lesssim u_0$.
If (for $\tau_\text{e}'<\tau_M$) the initial velocity $|\VV_0|\gtrsim u_0$ itself represents the maximal (persistent) velocity, 
the behavior of MSD is independent of the other parameters.

\section{Conclusions \label{sec_concl}}

In conclusion, we have explored an active Ornstein-Uhlenbeck particle (AOUP) with inertia and calculated
analytically various  dynamical correlation functions such as the mean-square displacement (MSD).
In particular, we extended  recent work \cite{inertial} by including the explicit dependence on the initial velocity
and by considering unstable inverted harmonic potentials, two coupled dumbbell-like particles
and the situation of a time-dependent mass. Different dynamical scaling regimes were identified including
power laws where the MSD scales in time $t$ with a power law $t^\alpha$.
Here, the dynamical scaling exponent can be
$\alpha\in\{0,1,2,3,4\}$. These scalings resemble results in other situations such as for an active Brownian particle (ABP)
in  a linear shear field \cite{Wittkowski_PRE_2012} or
 a disordered potential energy landscape \cite{breoni2020}.

In principle, our predictions can be tested in experiments on
macroscopic self-propelled  particles or mesoscopic particles
in a gaseous background. 
Examples from the inanimate macroscopic world include
vibration-driven granular particles \cite{NarayanRM2007,KudrolliLVT2008,DeseigneDC2010,GiomiHWM2013,WeberHDLDFC2013,KlotsaBHBS2015,PattersonFSJKGZPP2017,JunotBLD2017,Ramaswamy2017,DeblaisBGDVLBBK2018,DauchotD2019},
autorotating seeds and fruits \cite{RabaultFC2019,FauliRC2019}, 
camphor surfers \cite{LeoniPEENASA2020}, 
hexbug crawlers \cite{LeoniPEENASA2020}, 
trapped aerosols \cite{DiLeonardoRLPWGBM2007} and
mini-robots \cite{RubensteinCN2014,FujiwaraKI2014,TolbaAR2015,ZhakypovMHP2019,YangRCZ2019}.
Another system which has gained more recent attention are
complex plasmas consisting of mesoscopic charged dust particles \cite{MorfillI2009,SuetterlinEtAl2009,CoueedelNIZTM2010,ChaudhuriIKTM2011,IvlevBHDNL2015,NosenkoLKRZT2020,Petrov}. Furthermore,
 animals moving at intermediate Reynolds number exhibit inertial effects such as swimming organisms like
nematodes, brine shrimps or whirligig beetles \cite{Klotsa2019,beetlePREPRINT} and flying insects and birds \cite{TonerTu1995,Flocks1998,Chiappini2008,Bartussek2016,Mukundarajan2016,Bartussek2018,Attanasi2014}.
Since, at low Reynolds numbers, a passive particle in a sea of active particles was shown to be an excellent realization of
overdamped AOUP \cite{Maggi2014expAOUP, Maggi2017expAOUP}, one might expect that a macroscopic (inertial) passive particle
in a background of other active particles will realize an inertial AOUP but this conjecture needs to be tested.

For the future, the inertial AOUP model can be extended to more complex situations. Among those
is an inertial circle swimmer, a situation which has been explored for overdamped ABPs \cite{CABP,circular1,circular2}
and overdamped AOUPs \cite{CAOUP},
and motions under an external %
magnetic field \cite{magnet1,magnet2} or in non-inertial frames \cite{frame1,frame2}. Last the collective behavior
of many inertial active particles,
such as MIPS \cite{SumaGMO2014,ScholzRotor2018,PetrelliDCGS2018,MayyaNSHE2019,MandalLL2019,caprini_inertialABPSvelocity2021,omar2021_inertiaMIPS} or pattern formation in general \cite{AroldS2020},
is largely unexplored and our simple model may provide a stepping stone to access these fascinating phenomena.

\appendix

\section{Additional and full analytic results \label{app_calc}}

First, for a free particle, the general VACF 
\begin{align}
& V_\text{f}(t_1,t_2) =\VV_0^2\,{\rm e}^{-\frac{\gamma(t_1+t_2)}{m}}+\frac {2\gamma D}{m^2-\gamma^2\tau^2} \cr 
&\ \ \cdot  \bigg\{ \gamma\tau\left({\rm e}^{-\frac{t_1}{\tau}-\frac{\gamma t_2}{m}}+{\rm e}^{-\frac{t_2}{\tau}-\frac{\gamma t_1}{m}}-{\rm e}^{-\frac{1}{\tau}(t_1-t_2)}-{\rm e}^{-\frac{\gamma(t_1+t_2)}{m}}\right)\cr
& \ \ \ \ \ +m\left({\rm e}^{-\frac{\gamma}{m}(t_1-t_2)}-{\rm e}^{-\frac{\gamma}{m}(t_1+t_2)}\right)\bigg\}\,,
\label{eq_VACFfull}
\end{align}
 calculated according to Eq.~\eqref{eq_calcvacf} and given here for the case $t_1 \geq t_2$, 
 does not only depend on the absolute difference $|t_1-t_2|$ because the system is not in steady-state. 
 Taking the steady-state limit, $\lim_{t' \to \infty}V_\text{f}(t'+t,t')$ yields the result stated in Eq.~\eqref{eq_freevacf}.
The MSD, Eq.~\eqref{eq_MSDfreeFULL}, is found from inserting the VACF from Eq.~\eqref{eq_VACFfull} into Eq.~\eqref{eq_msdvacf}.
In the steady state, the expression for the MSD reduces to Eq.~\eqref{eq_MSDfreeFULLss},
which can be seen by inserting the stationary VACF, Eq.~\eqref{eq_freevacf}, into Eq.~\eqref{eq_msdvacf}.

    Second, the full MSD for an AOUP in a harmonic potential, given by Eq.~\eqref{eq_harmonicpot}, is given by
  \begin{widetext}
 \begin{align}
\Delta_\text{h}(t) =&\; \left[\left(\frac{\gamma \mathbf{r}_0}{2 m}+\mathbf{v}_0 \right)\frac{\sinh(\mu t)}{\mu}\,{\rm e}^{-\frac{\gamma t}{2 m}}+\mathbf{r}_0 \left(\cosh(\mu t)\, {\rm e}^{-\frac{\gamma t}{2 m}}-1 \right) \right]^2 \cr
&+\frac{2D\gamma}{k m (\gamma^2 - 4 k m) (m^2 + 2 k m \tau^2 - \gamma^2 \tau^2 + k^2 \tau^4)} \cr
&\cdot \biggl\{ - \gamma k \tau^2 (\gamma^2-4 k m)  \biggl[2 m \tau \cosh(\mu t) +  (2 m + \gamma \tau) \frac{\sinh(\mu t)}{\mu} \biggr] {\rm e}^{-\left(\frac{\gamma}{2 m} + \frac{1}{\tau}\right)t}\cr
&\ \ \ \ + m (m + \gamma \tau + k \tau^2) \biggl[\frac{\gamma}{m} (\gamma^2 - 4 k m) (\gamma \tau - m) \cosh(\mu t)\, \frac{\sinh(\mu t)}{\mu}\cr
&\ \ \ \ \ \ \ \ \ \ \ \ \ \ \ \ \ \ \ \ \ \ \ \ \ \ \ \ \  \ \ \  + 2 \gamma (\gamma^2 \tau - 2 k m \tau - \gamma m) \cosh^2(\mu t) - \gamma^3 \tau + \gamma^2 m + 4 k m^2 \biggr] {\rm e}^{-\frac{\gamma t}{m}} \nonumber\\
&\ \ \ \ + m (\gamma^2 - 4 k m) (m + \gamma \tau) (m - \gamma \tau + k \tau^2) \biggr\}\!\!\!\!\!\!\!
\end{align}
  \end{widetext}
with $\mu \coloneqq \frac{\sqrt{1 - 4 \frac{ \tau_m }{ \tau_k}\text{sgn}(k)}}{2 \tau_m} = \frac{\sqrt{\gamma^2 - 4 k m}}{2 m}$.


%
\end{document}